\documentclass[doublecol]{epl2} 

\usepackage[T1]{fontenc}
\usepackage{newtxtext}
\usepackage{newtxmath}

\usepackage{amsmath}
\usepackage{mathtools}
\usepackage{amsfonts}
\usepackage{dsfont}
\usepackage{braket}

\usepackage{color}
\usepackage{xcolor}
\usepackage[colorlinks,citecolor=blue,linkcolor=blue,urlcolor=blue]{hyperref}

\usepackage{graphicx}
\usepackage[all]{hypcap} 

\newcommand{\Tr}[1]{\text{Tr}\, #1}

\title{Unified theory of classical and quantum ergotropy}
\shorttitle{} 

\author{Michele Campisi}
\shortauthor{M. Campisi}

\institute{                    
  \inst{1} Istituto Nanoscienze-CNR, NEST Scuola Normale Superiore, 56127 Pisa, Italy
}

\abstract{
Quantifying the ergotropy (a.k.a. available energy), namely the maximal amount of energy that can be extracted from a thermally isolated system, is a central problem in quantum thermodynamics. Notably, the same problem has been long studied for classical systems as well, e.g. in plasma physics and astrophysics, where the basic principles for its solution are known for the case of collisionless fluids. Here we provide the general analytical expression of ergotropy of classical systems valid regardless of their size and the type of interparticle interactions, and show that it emerges as the classical limit of the quantum expression of ergotropy, for  quantum systems that are classically ergodic. We thus establish a unified theory of classical and quantum ergotropy, whose applicability ranges from atomic to galactic scale. Such unified theory is indispensable for studying the genuine quantum signatures of ergotropy: We show that the celebrated decomposition of quantum ergotropy into coherent ant inchoherent parts survives in the classical regime, indicating that coherences do not necessarily  reveal quantumness. The unified theory also allows to port tools and methods across the classical-quantum boundary to unlock the solution of standing problems. We apply this to swiftly solve the open problem of ergotropy extraction in the classical regime.
}

\begin{document}

\maketitle

What is the maximal amount of energy that can be extracted from a thermally isolated system by means of a cyclical external driving? And what is the driving that achieves such maximal energy extraction? 

These questions are in the limelight of current investigation in the field of quantum thermodynamics \cite{Gemmer09Book,Binder18Book,Deffner19Book,Campbell26QST11} where they naturally emerge in the study of quantum batteries \cite{Alicki13PRE87,Binder15NJP17,Ferraro18PRL120,Campaioli24RMP96,Ferraro26NRP}, quantum heat engines \cite{Quan07PRE76,Allahverdyan08PRE77,Allahverdyan13PRL111,Campisi15NJP17,Cangemi24PR1087} and dynamic cooling \cite{Allahverdyan10PRE81,Allahverdyan11PRE84,Park16inbook16,Oftelie24PRXQ5,Xuereb25PRXQ6}.
The concept of maximal energy extractable from a thermally isolated  system was first introduced, in the context of quantum theory, in 1976 in Ref.  \cite{Hatsopoulos76FP6b} where the expression ``adiabatic availability' was used to denote it. However the topic  took momentum only later with the  work of Allahverdyan \emph{et al.} \cite{Allahverdyan04EPL67}, who coined the expression ``ergotropy'' for the same quantity, and provided its analytical expression along with  the expression of the according "ergotropy extracting" unitary operator, see Eqs. (\ref{eq:optimal-U},\ref{eq:q-ergotropy}) below. 

Here we note that the very same quantity has been long studied for classical fluids as well, where it is known as ``available energy'' (often indicated by the symbol \AE).
The study of a fluid's available energy can be traced back to Lorenz who introduced it in atmospheric physics in 1955 \cite{Lorenz55TELLUS7} and has found applications ranging from 
plasma physics, \cite{Gardner63PF6,Dodin05PLA341,Helander17JPP83,Kolmes20PRE102,Qin25PRE11,Helander24JPP90,Mackenbach22PRL128}, to astrophysics \cite{Wiechen88MNRAS232,Lemou12IM187} and climate science \cite{Singh22RMP94}. In this context, available energy is a
fundamental tool for the study of fluid instabilities:  the higher the available energy the farther the system is from a stable equilibrium.

In the pioneering and elegantly succinct work of Ref. \cite{Gardner63PF6} Gardner pinpointed the fundamental principles that guide the calculation of the available energy of a collisionless fluid obeying Vlasov dynamics. Their application ranges from plasmas \cite{Dodin05PLA341,Helander17JPP83} to self-gravitating matter \cite{Wiechen88MNRAS232}. Here we note that Gardner's principles can be applied to the calculation of ergotropy of any classical (possibly many-body interacting) Hamiltonian system, whose phase space density evolution is Liouvillean. We show how they univocally  lead to the general analytical expression of ergotropy of a classical system, see Eqs. (\ref{eq:c-ergotropy},\ref{eq:c-ergotropy3}). Most importantly we show that, for quantum systems that are classically ergodic, such expression emerges as the classical limit of the celebrated quantum expression of ergotropy  \cite{Allahverdyan04EPL67}, Eq. (\ref{eq:q-ergotropy}). 

Thus we bridge and unify results obtained in distant subfields of physical investigation, such as plasma- and astro-phyiscs and quantum thermodynamics, while greatly extending their realm of applicability. 

The present unified theory of ergotropy provides the solid theoretical ground for unlocking open questions both in the classical and quantum realms.
For example a timely topic of quantum thermodynamic investigation concerns the identification of  the genuinely quantum features of non-equilibrium energy exchange processes \cite{Gemmer09Book,Binder18Book,Deffner19Book,Campbell26QST11}. Current expectation, based on the recently reported decomposition of quantum ergotropy into coherent and inchoherent parts \cite{Francica20PRL125}, is that coherences reveal its genuinely quantum features. Here we prove that the decomposition survives in the classical regime thus showing that coherences do not necessarily reveal quantum signatures of ergotropy. The main tool used to address this point is the classical dephasing operator, first introduced in \cite{Smith22entropy22} to study the impact of coherences  in the context of isothermal (rather than adiabatic) work extraction: In this sense the present work is complementary to that of  \cite{Smith22entropy22}.

Clearly, a unified theory also allows to port ideas and methods across the quantum-classical boundary. This can be used as a strategy to address open problems on both sides. For example we apply it to solving the pressing, yet unsolved, problem of finding a driving protocol that extracts the ergotropy in the classical case.

\section{Quick review of ergotropy of a quantum system }
Given  a quantum system with Hamiltonian $\hat{H}_0$ being in a state described by the density operator $\hat{\rho}_0$, we ask what is the unitary evolution operator  $\hat{U}$ that leads to the state $\hat \rho_1= \hat{U} \hat{\rho}_0 \hat{U}^\dagger$ of lowest energy expectation
\begin{align}
\breve{E}_q = \min_{\hat U} \Tr \hat{H}_0  \hat{U} \hat{\rho}_0 \hat{U}^\dagger\,.
\end{align}
For Hamiltonians with discrete spectrum acting on a Hilbert space of finite dimension $d$ the above minimum is reached by the  unitary operator \cite{Allahverdyan04EPL67} 
\begin{align}
\hat{U} = \sum_{k=1}^d  e^{-i \phi_k} |e_k\rangle \langle r_k|  \, ,
\label{eq:optimal-U}
\end{align}
where $\phi_k$ are irrelevant phases,
$\ket{e_k}$ ($\ket{r_k}$) are the eigenvectors of $\hat{H}_0$ ($\hat{\rho}_0$) relative to the according eigenvalues $e_k$ ($r_k$) ordered in non-decreasing (non-increasing) fashion:
\begin{align}
\hat{H}_0 &= \sum_k  e_k  |e_k\rangle \langle e_k|, \quad e_1 \leq e_2 \leq \dots \leq e_d \\
\hat{\rho}_0 &= \sum_k  r_k  |r_k\rangle \langle r_k|, \quad r_1 \geq r_2  \geq \dots \geq r_d \, .
\end{align}
Under the evolution $\hat{U}$ the system reaches the minimal energy state
\begin{align}
\hat{\rho}_1 = \hat{U} \hat{\rho}_0 \hat{U}^\dagger = \sum_k r_k  |e_k\rangle \langle e_k|
\label{eq:rho1-q}
\end{align}
and the quantum ergotropy, i.e., the difference between the initial energy expectation $E_0 = \Tr \hat{H}_0  \hat{\rho}_0$ and $\breve{E}_q$ amounts to
\begin{align}
\mathcal{E}_q = \sum_{j,k}  r_j (|\langle r_j |e_k \rangle |^2 - \delta_{jk} )e_k\,,
\label{eq:q-ergotropy}
\end{align}
where $\delta_{jk}$ denotes the Kronecker delta.

\section{Ergotropy of a classical system}
Consider a classical system with unperturbed Hamiltonian $H_0(\mathbf z)$ being initially in a statistical state described by the phase space distribution 
 $\rho_0(\mathbf z)$. Here $\mathbf z= (\mathbf p, \mathbf q)$ denotes a point in phase space. We ask what is the distribution $\rho_1(\mathbf z)$ featuring the smallest energy expectation among all those that can be connected to $\rho_0$ by a Hamiltonian flow. Two ingredients are sufficient to answer this question. First, as  formally demonstrated in \cite{Gorecki80LMP4,Daniels81JMP22}, each and all states of the form $\rho_0=g(H_0)$ with $g$ a non-increasing function are ``passive'' states, meaning are such that no energy can be further extracted therefrom. 
 This means that $\rho_1$ must be of the form:
\begin{align}
\rho_1(\mathbf z)=g(H_0(\mathbf{z})), \quad g(x)\leq g(y), \text{for } x >y
\label{eq:Gardner1}
\end{align}
Second, not only any Hamiltonian flow preserves the volume in phase space (Liouville theorem) but also any volume preserving map can be arbitrarily well approximated by some (possibly non-smooth) Hamiltonian flow (see Ref. \cite{Katok73MathUSSRIzv7} for a rigorous proof, or Appendix A for more intuitive, less formal, argument). This extends the optimisation space from the set of Hamiltonian maps, to the more easily treatable set of ``volume preserving" maps $\mathcal M:\mathbb{R}^{2s}\to\mathbb{R}^{2s}$. In particular this would imply that for any $r>0$, the volume of phase space where $\rho_1 >r$ must be equal to that where $\rho_0 >r$, that is:
\begin{align}
\int d\mathbf z \theta(\rho_0(\mathbf z)-r) = \int d\mathbf z \theta(\rho_1(\mathbf z)-r) 
\label{eq:Gardner2}
 \end{align}
where $\theta$ denotes Heaviside step function. To see that note that the set $A_0=\{\mathbf z \in \mathbb{R}^{2s}| \rho_0(\mathbf z)> r\}$ gets mapped onto $A_1=\{\mathbf z \in \mathbb{R}^{2s}|\rho_0(\mathcal M(\mathbf z))> r\}$. Since by definition $\rho_1(\mathbf z)= \rho_0(\mathcal M(\mathbf z))$, Eq. (\ref{eq:Gardner2}) expresses that the volume of $A_1$ is the same as that of $A_0$, for any $r>0$.

Eqs. (\ref{eq:Gardner1},\ref{eq:Gardner2}) where first noted by Gardner  \cite{Gardner63PF6} for the case of a collisionless plasma, for which $\rho_0$ is a single particle distribution obeying Vlasov equation. Their validity is however general: they hold for any (possibly many-body interacting) Hamiltonian system. Most importantly they univocally single-out the function $g$ that determines the passive companion, $\rho_1$, of $\rho_0$. To see that, let
\begin{align}
\Sigma(r)=  \int d\mathbf z \theta[\rho_0(\mathbf{z})-r]
\label{eq:R}
\end{align}
denote the volume of phase space where $\rho_0 > r$. Note that $\Sigma$ is a decreasing function. For simplicity, in the following we shall assume $\rho_0$ is continuous and without flat plateaus, in which case $\Sigma$ is continuous and strictly decreasing, hence invertible. We will comment below on the important physical meaning of the function $\Sigma^{-1}$. Meanwhile, let 
\begin{align}
\Omega_0(E) = \int d\mathbf z \theta[E-H_0(\mathbf{z})]
\label{eq:Omega_0}
\end{align}
denote  the volume of phase space where $H_0 \leq  E$\cite{Boltzmann85JRAM98,Hertz10AP338a,Hertz10AP338b,Einstein11AP34,Khinchin49Book,Campisi05SHPMP36,Campisi10AJP78,Dunkel14NATPHYS10,Campisi25Book}. 
For physical Hamiltonians,  $\Omega_0$ is a strictly increasing, hence invertible, function. We shall denote its inverse as $E_0(\cdot)\doteq \Omega_0^{-1}(\cdot)$, so that $E_0(\Phi)$ represents the energy of the iso$-H_0$ hypersurface that encloses the volume $\Phi$. In the following, for any generic phase function $F(\mathbf z)$ we shall call the iso-$F$ hypersurfaces, simply the ``eigensurfaces'' of $F$.
Combining Eqs. (\ref{eq:Gardner1},\ref{eq:Gardner2},\ref{eq:R},\ref{eq:Omega_0}) we get
\begin{align}
\Sigma(r) &= \int d\mathbf z \theta[g(H_0(\mathbf{z}))-r]\nonumber \\
&= \int d\mathbf z \theta[g^{-1}(r)- H_0(\mathbf{z})] = \Omega_0(g^{-1}(r)) \, ,
\end{align} 
that is $\Sigma= \Omega_0 \circ g^{-1}$. It follows that
$
g = \Sigma^{-1} \circ \Omega_0
$ \footnote{Note that since $\Sigma^{-1}$ is decreasing, and $\Omega_0$ is increasing, $g$ is decreasing as asuumed from the start.}. 
Combining with Eq. (\ref{eq:Gardner1}) we get:
\begin{align}
\rho_1(\mathbf z)=\Sigma^{-1}(\Omega_0(H_0(\mathbf z)))\, .
\label{eq:ground}
\end{align}
This explicit expression of the passive companion, $\rho_1$, of $\rho_0$, in Eq. (\ref{eq:ground}) is our first crucial result. It generalises similar ones obtained in plasma/astro-physics for collisionless systems \cite{Wiechen88MNRAS232,Helander17JPP83}. Among plasma physicists $\rho_1$ is known as the ``ground state'' associated to $\rho_0$ while astrophysicists call it its ``associated equilibrium'' or simply ``lowest energy state'' \cite{Wiechen88MNRAS232}. Mathematically speaking, $\rho_1$  would be  the "ergotropic rearrangement'' of $\rho_0$ 
\cite{Campisi26arXiv:2603.28388} that generalises the standard notion of ``symmetric decreasing rearrangement''  \cite{Hardy34Book,Lieb01Book}. 

Given $\rho_1$, Eq. (\ref{eq:ground}), the minimal energy reads:
\begin{align}\label{eq:Gardner-free-energy}
&\breve{E}_c =  \int d \mathbf{z} H_0(\mathbf{z})\rho_1(\mathbf z)
=  \int d \mathbf{z} H_0(\mathbf{z}) \Sigma^{-1}(\Omega_0(H_0(\mathbf z))) =\nonumber \\
& \int_0^\infty de\, \omega_0(e) e \Sigma^{-1}(\Omega_0(e))= \int_0^\infty d\Phi  E_0(\Phi) \Sigma^{-1}(\Phi)
\end{align}
where we first changed the integration variable from phase-space $\mathbf z$ to energy $e=H_0(\mathbf{z})$, using the density of states $\omega_0(e)= \Omega_0'(e)$, and then made the change of variable $e \rightarrow \Phi=\Omega_0(e) $. 
The ergotropy $\mathcal E_c $of a classical system is then:
\begin{align}
\mathcal E_c = \int d \mathbf{z} H_0(\mathbf{z}) \rho_0(\mathbf{z}) - \int_0^\infty d\Phi  E_0(\Phi) \Sigma^{-1}(\Phi) \, .
\label{eq:c-ergotropy}
\end{align}

Note that since $E_0$ has the units of energy, $\Sigma^{-1}$ has units of a probability density in ``phase volume''.
In fact it can be interpreted as the probability density, call it $P_1(\Phi)$, that the system is found on the eigensurface of $H_0$ that encloses the volume $\Phi$ provided it is distributed according to $\rho_1$.  To see that we express $P_1$ as the overlap of $\rho_1$ and the microcanonical state $\rho_\mu$ that ``lives'' on that eingensurface
\begin{align}
\rho_{\mu}(\mathbf z;\Phi)= \delta[\Phi-\Omega_0(H_0(\mathbf z))]
\end{align}
Note that here the microcanonical distribution is parametrised via the enclosed-volume $\Phi$ rather than the usual energy $E$, which makes it naturally normalised. Such unusual, yet convenient, parametrisation, which is crucial in this work, was first put forward in Ref. \cite{Campisi25Book}. 
As anticipated above, we have:
\begin{align}
P_1(\Phi)&\doteq \int d\mathbf z \rho_1 (\mathbf z) \delta[\Phi-\Omega_0(H_0(\mathbf z))]\label{eq:P1}\\
=\int d\mathbf z&  \Sigma^{-1}(\Omega_0(H_0(\mathbf z)))\delta[\Phi-\Omega_0(H_0(\mathbf z))]=\Sigma^{-1}(\Phi)  \, . \nonumber 
\end{align}
Introducing the probability density $P_0(\Phi)$ of finding the system on the eigensurface of $H_0$ that encloses the volume $\Phi$, provided it is distributed according to $\rho_0$
\begin{align}
P_0(\Phi) =  \int d\mathbf z \rho_0(\mathbf z) \delta[\Phi-\Omega_0(H_0(\mathbf z))]\, .
\end{align}
the ergotropy takes the symmetric and meaningful form:
\begin{align}
\mathcal E_c =  \int d \Phi [P_0(\Phi)-P_1(\Phi)] E_0 (\Phi)\,. \label{eq:c-ergotropy2}
\end{align}

Most remarkably $P_0(\Phi)$ is linked to the probability density $P_1$ as follows:
\begin{align}
&P_0(\Phi) \nonumber \\
&= \int d \Theta \int d\mathbf z \delta[\Phi-\Omega_0(H_0(\mathbf z))] \delta[\Theta-\Sigma(\rho_0(\mathbf z))]  \rho_0(\mathbf z)
\nonumber \\
&= \int d \Theta \int d\mathbf z \delta[\Phi-\Omega_0(H_0(\mathbf z))]  \delta[\Theta-\Sigma(\rho_0(\mathbf z))]  \Sigma^{-1}(\Theta) \nonumber \\
&=  \int d \Theta G[\Theta|\Phi] P_1(\Theta)\,, \label{eq:P0-P1}
\end{align}
where we first inserted the unit resolution, $1=\int d\Theta \delta (\Theta-y)$, and then introduced the overlap
\begin{align}
\hspace{-0.5mm}G[\Theta|\Phi] = \int d\mathbf z \delta[\Phi-\Omega_0(H_0(\mathbf z))]  \delta[\Theta-\Sigma(\rho_0(\mathbf z))] 
\label{eq:G}
\end{align}
between the microcanonical states having as support the eigensurfaces of $\rho_0$ and $H_0$ enclosing the volumes $\Theta$ and $\Phi$, respectively  \cite{Campisi25Book}.

Combining Eqs. (\ref{eq:c-ergotropy2}, \ref{eq:P0-P1}) we finally get
\begin{align}
\mathcal E_c =  \int d \Phi d\Theta P_1(\Theta) (G[\Theta|\Phi] - \delta[\Phi-\Theta] ) E_0 (\Phi)\,.
\label{eq:c-ergotropy3}
\end{align}
Note the formal analogy with the quantum formula (\ref{eq:q-ergotropy}) whereby the discrete principal quantum numbers $k,j$ are replaced by the continuous "enclosed volumes" variables $\Theta, \Phi$, and the quantum overlap of energy eigenstates is replaced by the classical overlaps of microcanonical states. 
The overlap $G$ can be understood as a special case of a classical propagator of a phase-volume density, which is the analogous of a quantum transition matrix. Such propagators were first discussed in Refs. \cite{Campisi08SHPMP39,Campisi08PRE78b}, and were  proved to be doubly-stochastic in Refs.
\cite{Deng17PRE95,Campisi25Book}. 

\section{Classical limit}
The  analogy between Eqs. (\ref{eq:q-ergotropy}, \ref{eq:c-ergotropy3}) is not just aesthetic, but it is substantial.
In order to see that consider the phase space formulation of quantum mechanics, based on the Wigner-Weyl transform  \cite{Schleich01Book,BalianBook1}. Note first of all that the Weyl representation $H^W_0(\mathbf z)$ of the operator $\hat H_0$  tends in the classical limit to its classical expressions $H_0(\mathbf z)$, where the position and momentum operators are replaced by real variables. Next note that the  Wigner function $W_{\hat \rho_0}(\mathbf z)$ of the initial preparation tends to a proper Liouville density $\rho_0(\mathbf z)$; similarly, the final state $W_{\hat \rho_1}(\mathbf z)$ tends to the Liouville density $\rho_1(\mathbf z)$. Also the latter is linked to the former by the Liouville dynamics because the Moyal dynamics dictating the evolution of the Wigner functions becomes Liouvillean in the classical limit \cite{Schleich01Book}.

Let's then consider the probability $r_k$. From Eq. (\ref{eq:rho1-q}) it is:
\begin{align}
r_k= \Tr \hat \rho_1 \ket{e_k}\bra{e_k} = \int d\mathbf z W_{\hat \rho_1}(\mathbf z) W_{ \ket{e_k}\bra{e_k}}(\mathbf z)
\label{eq:rk-wigner}
\end{align}
For classically ergodic quantum systems\footnote{That is, when the Weyl representation of the Hamilton operator tends, in the classical limit, to an ergodic Hamilton function in phase space}, the Wigner functions of the energy eigenstates tend to microcanonical classical states\footnote{See Sec. 3.5 of \cite{Schleich01Book}. This book discusses the emergence of the microcanonical distribution for simple 1D systems, but the derivation can be repeated for any ergodic system.}.  Then, assuming $\hat H_0$ is classically ergodic, the discrete set of the Wigner functions $W_{\ket{e_k}\bra{e_k}}(\mathbf z)$ indexed by the quantum number $k$, gets denser and denser thus approaching the continuous set of classical microcanonical distributions $\delta[\Phi-\Omega_0(H_0(\mathbf{z}))]$, indexed by the enclosed volume $\Phi$. Accordingly, the discrete sums become integrals and the expressions in Eq. (\ref{eq:rk-wigner}) tend to the quantities $P_1(\Phi)$ in Eq. (\ref{eq:P1}).
Similarly,  the eigenvalues
\begin{align}
e_k =  \Tr \hat H_0 \ket{e_k}\bra{e_k} = \int d\mathbf z H_0^W(\mathbf z) W_{\ket{e_k}\bra{e_k}}(\mathbf z)
\end{align}
go over the continuous function 
\begin{align}
E_0(\Phi)=\int d\mathbf{z}H_0(\mathbf{z})\delta(\Phi-\Omega_0(H_0(\mathbf{z})))
\end{align}
specifying the energy of each microcanonical eigenstate, labelled by the phase  volume enclosed by their support.

Finally, the quantum transition probabilities can be expressed as:
\begin{align}
p_{kj} =|\langle r_j |e_k \rangle |^2= \int d\mathbf z\,  W_{\ket{r_k}\bra{r_k}}(\mathbf z) W_{\ket{e_j}\bra{e_j}}(\mathbf z)
\end{align}
Interpreting the quantity  $f(\hat \rho_0)$ (with $f$ a generic strictly decreasing function), as a Hamilton operator (call it $H_1$) and assuming $H_1$ is classically ergodic too, the $p_{kj}$'s go,   in the classical limit, over the classical overlaps $G[\Theta|\Omega]$, Eq. (\ref{eq:G}). See also Ref. \cite{Jarzynski15PRX5} for explicit examples. All these observations demonstrate that, under the provision of ergodicity of $H_0,H_1$, the classical ergotropy, Eq. (\ref{eq:c-ergotropy}) emerges as the classical limit of the quantum ergotropy, Eq. (\ref{eq:q-ergotropy}).

\section{Extraction of ergotropy }
The above theory allows to apply methods and tools developed in the quantum case to the classical case and vice-versa. 
As an example we consider the problem of ergotropy extraction of a classical system. Notably, despite its practical interest, this problem has not been addressed so far in the literature.

Let us consider first the quantum problem, we ask what  time-dependent perturbation $\hat{V}(t)$  should one enact so that the according Hamiltonian 
$
\hat{H}(t)= \hat{H}_0 + \hat{V}(t)
$ 
would induce the ergotropy extraction unitary $\hat{U}$, Eq. (\ref{eq:optimal-U}). 
It is not difficult to see that this would be achieved by the following quench-adiabat (QA) protocol
 (i) instantaneously quench at $t=0$ to $\hat{H}_1 = f (\hat{\rho}_0) $ with $f$ some monotonous decreasing function; (ii) adiabatically return to $\hat H_0$. Under the provision of the adiabatic theorem   \cite{Messiah62Book} (namely if there are no level crossings during the adiabatic evolution) the eigenstates of $\hat H_1$, namely of $\hat \rho_0$, are dynamically mapped onto those of $\hat H_0$, and the fact that $f$ is monotonously decreasing ensures the right order is preserved so that the highest population state $\ket{r_1} $ gets mapped onto the ground state $\ket{e_1}$, the second most populated state $\ket{r_2}$ gets mapped onto the first excited state  $\ket{e_2}$, etc. , as per Eq. (\ref{eq:optimal-U}).

In classical mechanics one can do exactly the same QA protocol:
(i) instantaneously quench to 
$
H_1( \mathbf z) = f (\rho_0( \mathbf z))
$,
 for some monotonously decreasing function $f$, (ii) adiabatically return to $H_0(\mathbf z)$. 
Assuming that the motion induced by the ``frozen'' Hamiltonians $H(\mathbf z,t)$ is ergodic on their eigensurfaces \cite{Khinchin49Book,Campisi25Book} for all $t$'s in the QA protocol time span, the phase volume $\Omega_0$ is an adiabatic invariant \cite{Hertz10AP338b,Levi-Civita27Fermi,Ott79PRL42,Jarzynski92PRA46}. Consequently the QA protocol would dynamically map the eigensurfaces of $H_1(\mathbf{z})$  (that is the eigensurfaces of $\rho_0(\mathbf z)$) onto the eigensurfaces of $H_0(\mathbf z)$ that enclose same phase volumes thus realising the Gardner prescriptions. 

On a more intuitive level note that, just like in the quantum scenario, with the first step you lock the system in a passive state relative to $H_1$, and with the second step you adiabatically map it onto the passive state relative to $H_0$. This is well  illustrated by the example below.

\begin{figure}[t]
\includegraphics[width= 0.24\linewidth]{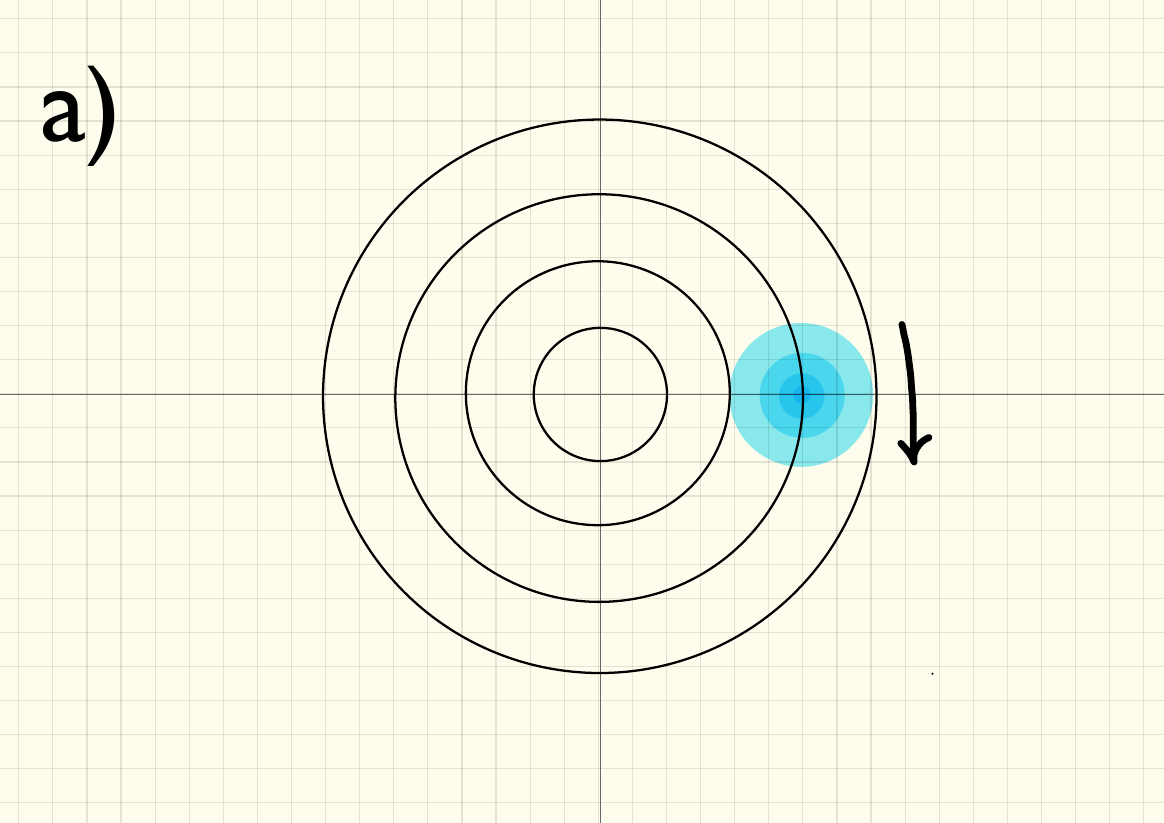}
\includegraphics[width= 0.24\linewidth]{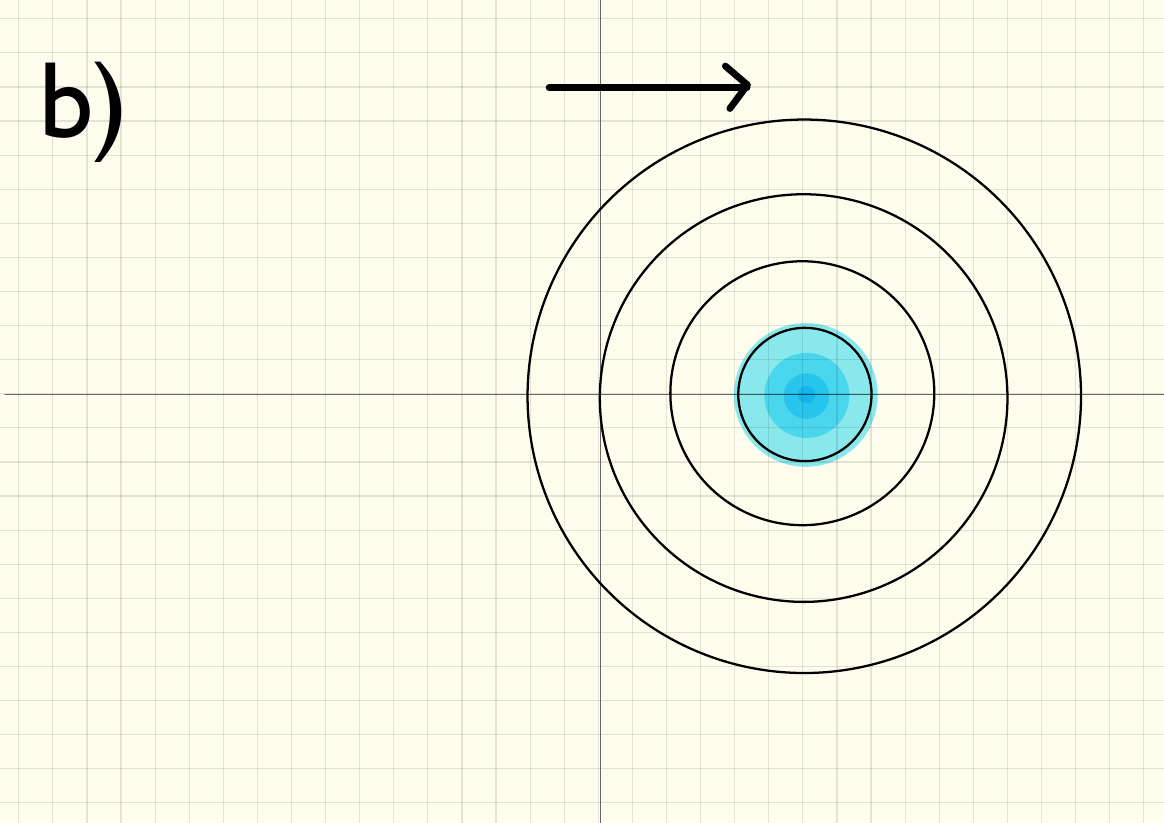}
\includegraphics[width= 0.24\linewidth]{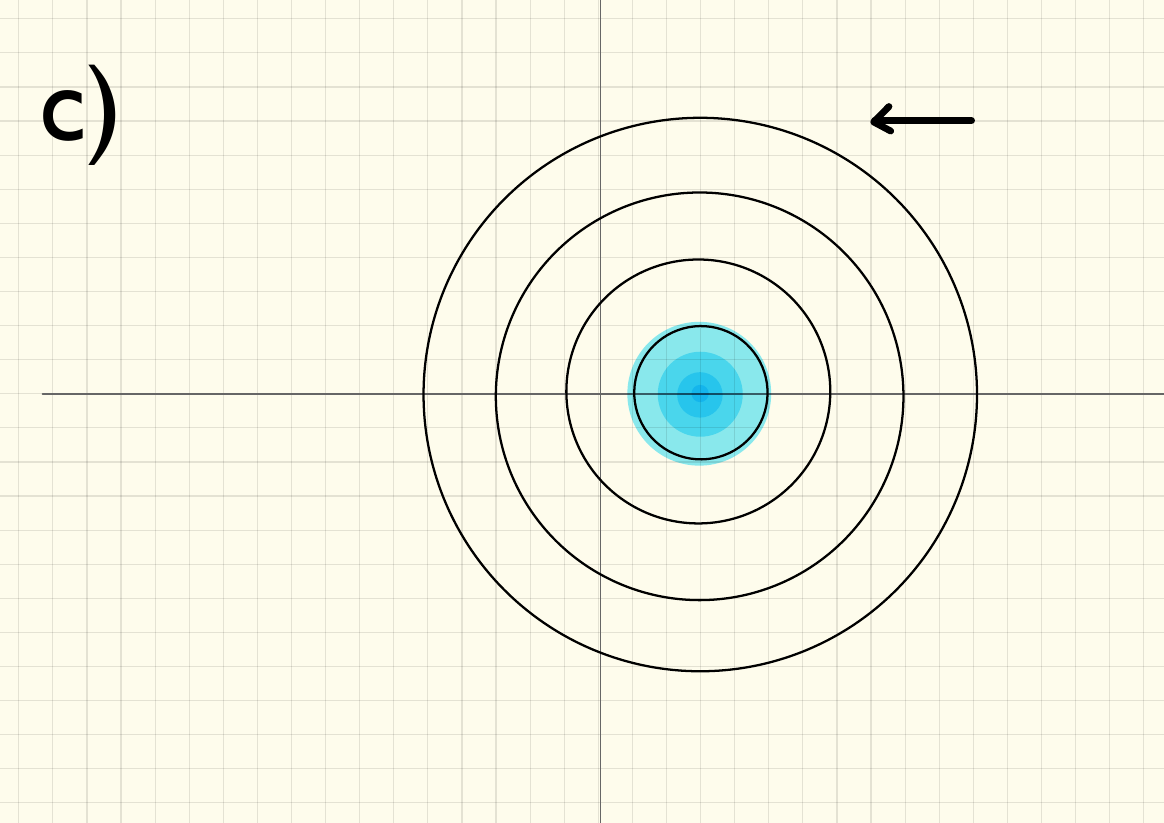}
\includegraphics[width= 0.24\linewidth]{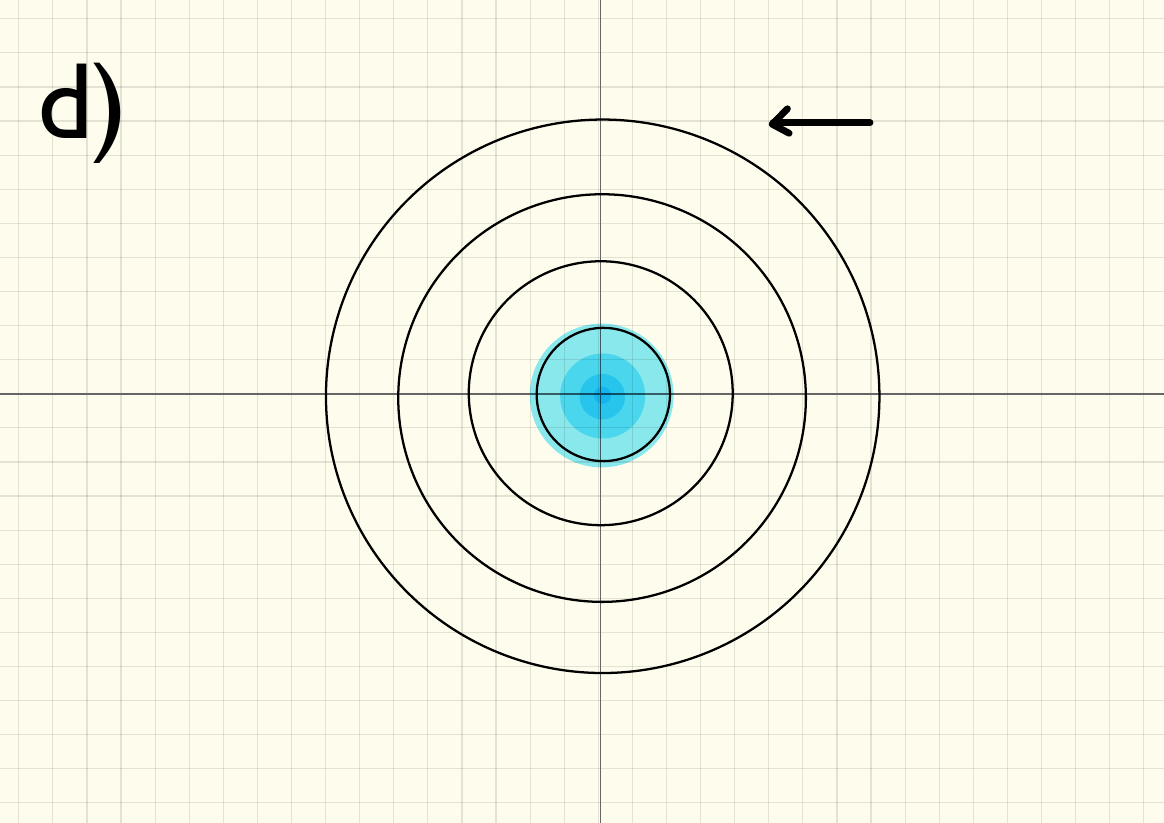}
\caption{Panels a-d): Phase space sketch of extraction of ergotropy from a non-stationary Gaussian state, Eq. (\ref{eq:HOgaussianstate}) of a harmonic oscillator. a) The phase density rotates under the unperturbed dynamics. b) At $t=0$ the harmonic potential gets instantaneously displaced so as to lock the state. c,d) In the time span $[0,\tau \gg1]$ the potential is slowly returned to its initial position.
}
\label{fig:coherent} 
\end{figure}
Consider a 1D Harmonic oscillator,
$H_0(q,p)= {p^2}/{2} + {q^2}/{2}$, 
being at $t=0$ in the Gaussian state
\begin{align}
\rho_0(q,p)= \frac{1}{2 \pi \sigma} \exp\left [- \frac{(q-q_0)^2+p^2}{2 \sigma} \right]\,.
\label{eq:HOgaussianstate}
\end{align}
See the sketch in Fig. \ref{fig:coherent}a). The state is clearly non-stationary (because $\{H_0,\rho_0\}\neq 0$) and would evolve by rotating around the phase space origin at angular frequency $\omega=1$, without changing shape, if unperturbed.
By suddenly displacing the harmonic potential  minimum at  $q=q_0$ blocks the state at the bottom of the displaced harmonic potential: the state is now stationary (in fact passive) relative to the new Hamiltonian 
\begin{align}
H_1(q,p)= \frac{p^2}{2} + \frac{(q-q_0)^2}{2}  = f( \rho_0(q,p))\,,
\label{eq:H1-example}
\end{align}
where $f(x)= -\sigma\ln(2\pi\sigma x)$ is a decreasing function, see Fig. \ref{fig:coherent}b).
By adiabatically moving the harmonic potential minimum back to the origin of the $q$-axis, the state evolves onto
\begin{align}
{\rho_1}(q,p)= \frac{1}{2 \pi \sigma}  \exp\left[- \frac{(q^2+p^2)}{2 \sigma} \right] =  \frac{e^{-H_0(q,p)/\sigma}}{2 \pi \sigma}\,,
\end{align}
because each instantaneous Hamiltonian visited by the protocol is ergodic, see Fig. \ref{fig:coherent}c,d). Since the state is passive relative to $H_0$ no more energy can be extracted therefrom.

\section{Decomopsition of ergotropy in coherent and incoherent parts}
The above unified theory of ergotropy is an indispensable tool for assessing the genuine quantum signatures of ergotropy. 
Ref.  \cite{Francica20PRL125} has highlighted that in the quantum case, the ergotropy splits into a coherent and an incoherent part. The standard expectation is that the coherent part reveals the quantum fingerprints of energy extraction in the quantum regime. Here we show that, contrary to that expectation, coherent contributions survive the quantum to classical transition, as the decomposition into a coherent and an incoherent part remains unaltered in the classical case.

The main tool that is needed for this analysis is the classical ``dephasing'' operator $\mathcal D$ relative to the Hamiltonian $H_0$
\begin{align}
\mathcal D[\rho](\mathbf z) = \int d \mathbf z'\, \rho( \mathbf z') \delta [\Omega_0(H_0( \mathbf z'))- \Omega_0(H_0( \mathbf z))]\, .
\label{eq:dephasing-op-c}
\end{align}
$\mathcal D$ homogenises a classical distribution $\rho$ over the eigensurfaces of $H_0$, thus rendering it ``diagonal'' with respect to $H_0$ (i.e., $\{\mathcal D[\rho],H_0\}=0$, where $\{\cdot,\cdot\}$ denotes Poisson brakets) while preserving its energy\footnote{The classical dephasing operator was first introduced in Ref. \cite{Smith22entropy22} where it was expressed in the energy parametrisation. Here we express it in the equivalent enclosed-volume representation, which we find more convenient.
}.
Exactly like in the quantum case \cite{Francica20PRL125}, we have
\begin{align}
 \mathcal E_c[\rho] =\mathcal E_c^c[\rho] + \mathcal E_c^i[\rho] \label{eq:split}
\end{align}
where 
\begin{align}
&\mathcal E_c^c[\rho] = \beta^{-1} (C[\rho] + D[\mathcal P[\mathcal D[\rho]] || \rho_\beta] - D[\mathcal P[\rho]|| \rho_\beta] )\label{eq:coherent}\\
&\mathcal E_c^i[\rho] =\mathcal E_c[\mathcal D[\rho]]  \label{eq:incoherent}
\end{align}
denote the coherent and incoherent parts, respectively. Here, in full analogy with the quantum case $\mathcal P[\rho]$ denotes the passive companion of $\rho$, $\rho_\beta = e^{-\beta H_0}/Z$ is a thermal state, $D[\cdot||\cdot]$ denotes the Kullback-Leibler divergence, and 
$
C[\rho] = D[\rho || \mathcal D[\rho]]  \label{eq:cocherence}
$
quantifies the ``coherence'' of $\rho$, that is how much it is distinguisgable from its dephased companion 
$\mathcal D[\rho]$.
Equations (\ref{eq:split}-\ref{eq:incoherent}) read exactly like their quantum counterparts where $D$ denotes the quantum Kullback-Leibler divergence, and $\mathcal D$ is the quantum dephasing operator relative to $\hat H_0$ \cite{Francica20PRL125}. The proof proceeds in a way that is fully analogous to the quantum case, see Appendix B for details. 

Note that both in quantum and classical cases lack of commutation between the state $\rho_0$ and the Hamiltonian $H_0$, either in the sense of quantum commutators or of Poisson brackets, signals that the state $\rho_0$ is not a function of $H_0$ hence it is not passive and its ergotropy is not null. Such lack of commutation, in both cases is associated to coherences, which in the classical case can be visualised as inhomogeneities of the phase distributions over the energy eigensurfaces \cite{Smith22entropy22}.

\section{Discussion}
We have developed a unified theory of classical and quantum ergotropy, a.k.a available energy. The theory bridges results obtained in distant parts of the  literature, i.e., plasma and astro-physics and quantum thermodynamics, by establishing a common dictionary and unifying them, while greatly extending their realm of applicability.

On the classical side the theory applies to any system obeying Liouville dynamics, regardless of the dimension $d$ of its phase space. This includes collisionless plasmas ($d=6$), single particle systems in D dimensions   (e.g., the harmonic oscillator of the above exameple $d=2$), but also many-body interacting systems ($d=6N$), like interacting gases and self-gravitating matter. Similarly, on the quantum side the theory applies to any quantum system obeying the quantum Liouville dynamics. Due to the classical limit, we can safely state that the present theory applies at scales ranging from atomic to galactic, form single particles, to many particles aggregates. 

We remark that our derivation of the classical limit relies on the ergodic hypothesis. Further investigations are needed to study the non-ergodic case. In this regard it is worth remarking that, roughly speaking, $H_1$ is ergodic if all its hypersurfaces are simply connected and the energy is the only first integral of motion. However these properties are fixed once and for all by the phase portrait of $\rho_0$ regardless of the shape of $f$. 

We also note that we have assumed $\Sigma$ is continuous and strictly decrasing, which excludes the case when  $\Sigma$ has plateaus and/or jumps. This case is treated in Ref. \cite{Campisi26arXiv:2603.28388}.

Thanks to the unified picture we were able to address open  questions existing both in the quantum and the classical realms. On the classical side, an open  pressing question is how to practically realise the time dependent perturbation that does indeed extract the ergotropy. By porting the simple quantum solution across the quantum-classical boundary, we swiftly formulate the formal solution of this open problem. The solution, the QA protocol, is extremely simple and intuitive in hind-sight, but yet never reported so far. We remark that its practical implementation may be greatly hindered, depending on the shape of $\rho_0$; e.g.,  when the eigensurfaces of $H_0$ and $\rho_0$ are not topologically equivalent, in which case it is impossible to link them smoothly and adiabatically. We leave a discussion of these challenging problems to future studies while anticipating that they could be solved by combining the present solution with known phase-space methods such as those that dynamically swap  regions of phase space nested one into the other \cite{Vaikuntanathan11PRE83,Holtzman25PRL134}. Despite these difficulties, our results form the basis for the systematic study of the problem of energy extraction in the classical regime.

On the quantum side an issue that is in the limelight of timely investigation concerns the quantification of genuinely quantum signatures associated to non-equilibrium energy extraction processes. The present unified theory is indispensable for a systematic study of this problem. Here, for example, we have highlighted that coherences (which in classical systems are revealed as inhomogeneities of the phase distribution  \cite{Smith22entropy22}) are a source of ergotropy both in the classical and quantum cases. Furthermore we showed that  the recently reported decomposition of ergotropy into coherent and incoherent components, which was suggested to be revealing of the quantum fingerprints of ergotropy \cite{Francica20PRL125}, survives in the classical regime. This indicates that, contrary to a widespread expectation, energy coherences do not necessarily constitute a genuine quantum thermodynamic advantage,  which corroborates the view expressed in  Ref. \cite{Smith22entropy22}. Most likely, quantum fingerprints should be looked among the genuinely quantum aspects of coherence, e.g., quantum entanglement, as suggested in Ref. \cite{Touil22JPA55}.
The issue remains open but now we have  powerful new tools for further systematic investigations. 

\acknowledgments
The author thanks Frank Ernesto Quintela Rodriguez for pointing out the plasma physics literature and Vittorio Giovannetti for pointing out Ref. \cite{Gorecki80LMP4}.
Discussions with Vasco Cavina, Marcus Bonança and Marcus Huber are gratefully acknowledged. The material presented here is based on lectures delivered at the University of Pisa, Italy,  in the academic year 2024/25.

\section{Appendix A}
A volume preserving map $\mathcal M: \mathbb R^{2s} \to \mathbb R^{2s}$ can be approximated to any wanted degree by a permutation of hypercubic phase space cells. The smaller the cells the better the approximation. The question is then, whether a permutation of hypercubic cells can be implemented by some Hamiltonian flow. Noting that any permutation can be decomposed in a finite sequence of swaps of adjacent cells, the question is whether a swap of two hypercubic cells can be implemented by some Hamiltonian flow. In 1D the answer is yes: consider the non-smooth Hamiltonian
\begin{align}
H(q,p)= \left\{\begin{array}{cc}
v |p|, &  (q,p) \in R\\
E, &  \text{otherwise}
\end{array}\right.
\end{align}
where $R$ is the rectangle $R=[-L/2,L/2] \times [-L,L]$, and $E > vL$. Inside $R$ the trajectories are rectangles. Their upper(lower) horizontal side is traversed at speed $+v(-v)$, while their vertical sides are traversed ``instantaneously" (just like it happens with the standard particle in box with reflecting walls). Outside $R$ all remains still. Thus, in the time $t=L/v$, the Hamiltonian flow swaps the square $[-L/2,L/2]\times [0,L]$ with its bottom neighbour $[-L/2,L/2]\times [-L,0]$, while leaving the rest of phase space unaltered. 
Similarly, for  $R=[-L,L] \times [-L/2,L/2]$ and $t=2L/v$ the Hamiltonian flow swaps the square $[-L,0]\times [-L/2,L/2]$ with its right neighbour $[0,L]\times [-L/20,L/2]$.
By translating the Hamiltonian one can then swap any two adjacent squares. The generalization to $s$ dimensions is straightforward.  Thus by concatenating swap-flows of the type above one can obtain a (non-smooth) Hamiltonian flow that implements any wanted permutation of phase-space cells and so approximate, to any wanted degree, any volume preserving map. 

\section{Appendix B}

To prove Eq. (\ref{eq:split}) we first note that
\begin{align}
&\int d\mathbf z \mathcal D[\rho](\mathbf z) f(H_0(\mathbf z)) \nonumber \\
&= \int d\mathbf z \int d \mathbf z'\, \rho(\mathbf z)  \delta[\Omega_0(H_0(\mathbf z'))-\Omega_0(H_0(\mathbf z))]f(H_0(\mathbf z)) \nonumber \\
&= \int d\mathbf z \int de\, \omega_0(e) \rho(\mathbf z)  \delta[\Omega_0(e)-\Omega_0(H_0(\mathbf z))] f(H_0(\mathbf z)) \nonumber \\
&= \int d\mathbf z \int d\Phi\, \rho(\mathbf z)  \delta[\Phi-\Omega_0(H_0(\mathbf z))] f(H_0(\mathbf z)) \nonumber \\
&=\int d\mathbf z\, \rho(\mathbf z) f(H_0(\mathbf z))\,, \label{q:int-rho-f-H}
\end{align}
where in the second line we have used the definition (\ref{eq:dephasing-op-c}), we then did the changes of variables $\mathbf z'\to e \to \Phi$, as, in Eq. (\ref{eq:Gardner-free-energy}), and finally used $\int d\Phi \delta(\Phi-y)=1$.
As a consequence of Eq. (\ref{q:int-rho-f-H}) we have:
\begin{align}
C[\rho] &= D[\rho|| \mathcal D[\rho]] = \int d\mathbf z \rho \ln \rho - \int d \mathbf z\, \rho \ln \mathcal D[\rho] \nonumber \\
&= \int d\mathbf z \rho \ln \rho - \int d \mathbf z\, \mathcal D[\rho] \ln \mathcal D[\rho]\, \nonumber \\
&=  \mathcal{H}[\mathcal D[\rho]] -\mathcal{H}[\rho]
\end{align} 
where $
\mathcal{H}[\sigma] =- \int d\mathbf z\, \sigma(\mathbf z) \ln \sigma(\mathbf z)
$ denotes the classical information. Using the well known formula:
\begin{align}
D[\sigma||\rho_\beta] = \beta \int d\mathbf z (\sigma(\mathbf z)-\rho_\beta(\mathbf z)) H_0(\mathbf z)
- \mathcal{H}[\rho]+\mathcal{H}[\rho_\beta]\,, \nonumber
\end{align}
where $\rho_\beta=e^{-\beta H_0(\mathbf z)}/{Z_0}$ is a thermal state, we have:
\begin{align}
\beta \mathcal E_c^c &= \beta (\mathcal E_c - \mathcal E_c^i) = \beta \int d\mathbf z H_0(\mathcal P[\mathcal D[\rho]]-\mathcal P[\rho]) \nonumber \\
&=\beta \int d\mathbf z H_0(\mathcal P[\mathcal D[\rho]]-\rho_\beta)- \beta \int d\mathbf z H_0(\mathcal P[\rho]-\rho_\beta) \nonumber \\
&= D[\mathcal P[\mathcal D[\rho]]||\rho_\beta] +\mathcal H[\mathcal P[\mathcal D[\rho]]]-\mathcal H[\rho_\beta]\nonumber \\
&\phantom{xxx}- D[\mathcal P[\rho]||\rho_\beta] -\mathcal H[\mathcal P[\rho]]+\mathcal H[\rho_\beta]\,. \nonumber
\end{align}
Equation (\ref{eq:coherent}) follows by noticing that $\mathcal H[\mathcal P[\rho]]=\mathcal H[\mathcal \rho]$,  $\mathcal H[\mathcal P[\mathcal D[\rho]]]=\mathcal H[\mathcal D[\rho]]$,  because, by construction, any state (be it $\rho$ or $\mathcal D[\rho]$), is linked to its passive companion ($\mathcal P[\rho]$ or  $\mathcal P[\mathcal D[\rho]]$, respectively) by a volume preserving map, and it is a well known fact that the classical information is invariant under such transformations.

\bibliographystyle{eplbib}

\end{document}